\documentclass{aastex}
\usepackage{spr-astr-addons}
\usepackage{url}
\usepackage{natbib}

\RequirePackage{color}

\newcommand{\emaila}{nadine.nettelmann@uni-rostock.de}
\newcommand{\sect}{\S$\:$}
\newcommand{\ME}{\:M_{\oplus}}
\newcommand{\RE}{\:R_{\oplus}}
\newcommand{\fig}{Fig.$\:$}

\begin{document}

\title{Predictions on the core mass of Jupiter and of giant planets in general\\
{\small\rm Accepted for publication in Astrophysics \& Space Science}}
\shorttitle{On the core mass of giant planets}
\shortauthors{N.~Nettelmann}

\author{Nadine Nettelmann}
\affil{University of California, Santa Cruz, Dept.~of Astronomy and Astrophysics, CA 95064, USA}
\affil{Universit\"at Rostock, Institut f\"ur Physik, D-18051 Rostock, Germany}
\email{\emaila}


\begin{abstract}
More than 80 giant planets are known by mass and radius. Their interior structure 
in terms of core mass, number of layers, and composition however is still poorly known. 
An overview is presented about the core mass $M_{core}$ and envelope mass of metals 
$M_Z$ in Jupiter as predicted by various equations of state. 
It is argued that the uncertainty about the \emph{true} H/He EOS in a pressure regime 
where the gravitational moments $J_2$ and $J_4$ are most sensitive, i.e.~between
0.5 and 4 Mbar, is in part responsible for the broad range $M_{core}=0-18\ME$,
$M_Z=0-38\ME$, and $M_{core}+M_Z=14-38\ME$ currently offered for Jupiter. We then 
compare the Jupiter models obtained when we only match $J_2$ with the range of 
solutions for the exoplanet GJ$\:$436b, when we match an assumed tidal Love number 
$k_2$ value.
\end{abstract}


\section{Introduction}

Jupiter is the best studied giant planet. Its oblate shape and the corresponding elliptic
gravity field deformation have been measured within an accuracy of $10^{-4}$. The 
uncertainty of its mass reflects the uncertainty of the gravitational constant $G$, and 
even its atmospheric He abundance is known within 3\% (see~\citealt{GuiGau07} for an overview). 
However, despite several spacecraft missions to Jupiter and observational data sampling over 
several centuries, we even do not know whether Jupiter possesses a core or not
~\citep{SauGui04}, and if heavy elements in the interior are distributed homogeneously. 
These are fundamental interior structure properties.
If we are to describe the interior of Jupiter and other giant planets, we shape our
imagination of \emph{interior structure} into terms such as \emph{core mass} and 
\emph{envelope metallicity} and apply the simplest possible structure type that
can quantitively answer our questions. For instance, the assumption of a completely 
homogeneous interior is sufficient to explain the observed mass and radius of a 
extrasolar giant planet, and an occasional additional constraint can be satisfied by
assuming the presence of a core within a two-layer approach. For solar giant planets however,
a two-layer approach is not consistent with available observational constraints when non-empirical
equations of state are applied (e.g.;~see \citet{Chabrier+92} for Jupiter and Saturn, and
\citet{Helled+10} for Uranus and Neptune), and hence we here model the interior by assuming 
three layers. This article is not about the difficulties we run into by relying on this approach. 
It is about the resulting core mass and metallicity of Jupiter (\sect\ref{ssec:Jup_McMZ}), 
where we repeat some of the analyis published in \citet{FN10} (hereafter FN10). We then 
continue in \sect\ref{ssec:Jup_ZZ} to investigate the specific open question if the 
assumption of an homogeneous heavy element dstribution can be subject to free choice 
in the case of recently published ab-initio data based Jupiter models, as it has been 
suggested in~\citet{MH09}.
In (\sect\ref{sec:J2k2}) we show the degeneracy of solutions raised by ignoring the 
octupole moment of the gravity field, $J_4$ equivalent to knowing the Love number $k_2$ 
of an extrasolar planet, but not $k_4$ .

\section{Modeling giant planets in isolation}\label{sec:methods}

\subsection{Observational constraints}

Interior models should reproduce the observed parameters. These are, first of all, 
mass $M_p$ and radius $R_p$ of the planet. For Jupiter, Saturn, and Uranus, the radius 
is the observable equatorial radius $R_{eq}$ taken at the 1~bar pressure level. Below that 
the interior is adiabatic. For close-in extrasolar giant planets, the observed 
transit radius occurs at 4$-$6 orders of magnitude lower pressures than the onset of the 
adiabatic region, and a proper outer boundary condition has to be provided by a model 
atmosphere. Calculations for particular giant planets predict the radiative/convective 
layer boundary deep in the atmosphere between 100 and 1000 bars~\citep{Showman+08}.
The mean helium abundance should equal that of the protostellar cloud where the 
planet formed from. However, the atmospheres of Jupiter and Saturn are depleted in 
helium compared to the solar value $Y=0.270\pm 0.005$ \citep{Bahcall+95}, which motivates 
the introduction of a He-poor outer envelope and a He-rich inner envelope. The 
metallicity $Z$ in the atmospheres of the solar planets is enhanced above the solar value 
$Z_{\odot}=0.015$ \citep{Lodders03} as derived from abundance measurements of single 
species (e.g.;~C, N, S, Ar, Xe). Solar planets are fast rotators, which gives rise to a 
gravity field deformation as expressed by the gravitational moments $J_2$, $J_4$, and 
$J_6$. Equivalent constraints for extrasolar planets are tidal Love numbers $k_{2n}$. 
Parameters related to thermal or orbital evolution can give additional constraints but 
are not addressed in this article.

\subsection{The Three-layer structure assumption}

For Jupiter and Saturn, the simplest structure type that is consistent with the 
constraints described above has three layers: a core of rocks and/or ices, and two 
envelopes that are convective, adiabatic, and homogeneous, but differ in the mass fraction
of helium $(Y)$ and metals $(Z)$. We choose the metallicities $Z_1$ and $Z_2$\footnote{Index 1 
refers to the outer envelope, and index 2 to the inner envelope.} 
to adjust J$_2$ and $J_4$. Often~(e.g;~\citealp{Gui99,Hori+08}) but not always~\citep{Militzer+08}
models that reproduce Jupiter's $J_2$ and $J_4$ are found to reproduce also $J_6$ within the 
obervational error bar. 
Physical reasons for a discontinuity in He or metals can be phase transitions, 
H/He phase separation, and in particular for metals the process of planet formation, 
but its location is essentially unconstrained, and hence the transition pressure 
$P_{1-2}$ between the envelopes a free parameter. In the absence of these particular
physical processes, gaseous planets are likely to have a homogenous envelope because 
of convection. If however convection is inhibited as suspected for Uranus~\citep{Podolak+91}
a series of double-diffusive layers can develop creating a large-scale compositional gradient.
Such complicated models are not considered here.

\subsection{Quasi-adiabat and equation of state}

At layer boundaries, pressure $P$ and temperature $T$ are required to transit 
continously. The $P-T-\rho$ profile, where $\rho$ is mass density, along the 
quasi-adiabat depends on the He abundances and metallicities chosen, 
and on the equations of state (EOS) of the underlying materials allowed for. 
If the envelopes differ in He mass fraction $(Y_1<Y_2)$ or metallicity $(Z_1\not=Z_2)$, 
then entropy and density transit discontinously, so we call the internal $P-T$ profile 
\emph{quasi}-adiabatic. At a layer boundary, a boundary layer can develop with 
change in temperature across or conductive heat transport, causing a warmer interior at 
higher entropy. The effect of a boundary layer has been included by \cite{ForHub03} 
for the inhomogeneous evolution of Saturn; but not yet in structure calculations. 
Since the gravity data probe the internal $P-\rho$ relation, the expected influence 
of a warmer interior on the structure would be a higher deep envelope metallicity.
 
It is convenient to represent elements heavier than He (\emph{metals}) 
by an equation of state of water \citep[e.g.;][]{ForHub03,SauGui04} as 
also supported by planet formation theory \citep{Helled+08}, and to assume the core be 
made of rocks \citep[e.g.;][]{ForHub03,SauGui04,N-Jupiter+08} or of water ice 
\citep[e.g.;][]{SauGui04}. Jupiter-sized giant planets are \emph{giant in size} because 
they are predominantly composed of the light elements H and He.   
For details of the EOS, see \citet{SauGui04,N-Jupiter+08}, FN10, and~\citet{Militzer+08}.

\subsection{Calculation of core mass and metallicity}

In order to obtain core mass and metallicity we integrate the 1-dimensional equations
of mass conservation, $dm/dl=4\pi l^2\rho(l)$, and of hydrostatic equilibrium, 
\begin{equation}\label{eq:hydro}
	\frac{1}{\rho(l)}\:\frac{dP}{dl}=\frac{dU}{dl}\:,
\end{equation}
along the quasi-adiabat for given outer boundary conditions and abundances of 
helium and metals. The coordinate $l$ parametrizes surfaces of constant total potential $U$. 
In spherical symmetry, i.e.~in the absence of rotation, $l$ equals the radial coordinate
$r$. The most often theory used to calculate the gravity field deformation 
according to shape deformation is the \emph{Theory of Figures} by \citet{ZT78}, 
but alternative theories can be applied as well \citep[e.g.;][]{HubbBuch84}.
The gravity field deformation on or exterior to the surface ($\rho(P=1\rm bar)\approx 0$)
can be described by multipole expansion into spherical harmonics, where the expansion
coefficients are the gravitational moments 
\begin{equation}
	J_{2n} = -\frac{1}{M_pR_{eq}^{2n}}\int_Vd^3r\rho(r,\theta)r^{2n}P_{2n}(\cos\theta)\:.
\end{equation}  
They are intergrals over the internal mass density within the volume $V$ of the planet.
Therefore, we can use the metallicities $Z_1$ and $Z_2$, which affect $\rho(l)$ 
along the quasi-adiabat, to adjust the moments $J_2$ and $J_4$. In general 
$Z_1\not=Z_2$. The core mass $M_{core}$ is then used to satisfy the inner boundary
condition $m(l=0)=0$ for given $M_p(R_p)$.

\section{Jupiter}\label{sec:Jupiter}

\subsection{Predicted core mass and envelope mass of metals}
\label{ssec:Jup_McMZ}

\begin{figure}[htb]
\plotone{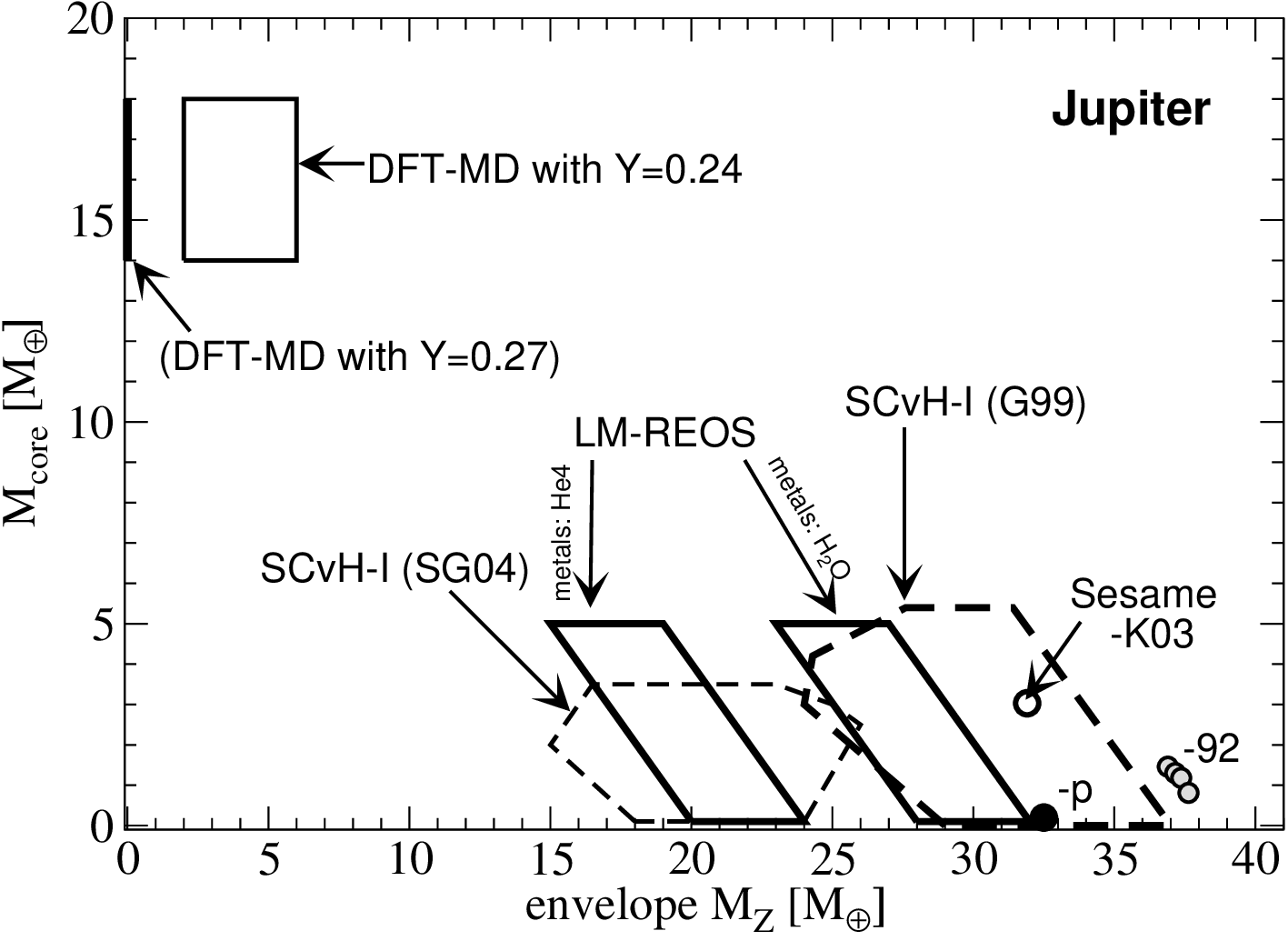}
\caption{\label{fig:Jup_McMZ}
Model results for the core mass ($M_{core}$) and mass of metals ($M_Z$) in the 
envelope(s) of Jupiter. Each polygon embraces the solutions found using a 
particular set of equations of state as labeled, for the components H, He, and metals. 
Models using SCvH-i (SG04) and Sesame-p assume $Z_1=Z_2$ and $Y_1<\bar{Y}<Y_1$ 
and are taken from~\citet{SauGui04}, SCvH-i (3L) based models have $Z_1\not=Z_2$~\citep{Gui99}, 
the Sesame-K04 model has $Z_1<Z_2$ \citep{KTS04-1} as have LM-REOS  based models 
(\citealp[updated from][]{N-Jupiter+08}), and the DFT-MD labeled model is a two-layer 
model with $Y_1=Y=0.24$ throughout the envelope. Adjusting this model for the protosolar 
mean value $Y=0.27$ by replacing $9\ME$ of metals by He yields a metal free envelope as
indicated. This figure shows $0\leq M_{core}\leq 18\ME$, $0\leq M_Z\leq 38\ME$, and
$14\leq M_{core}+M_Z\leq 38$, meaning that Jupiter's interior structure is not well known 
today and that models depend on the equation of state preferred.}
\end{figure}

Figure \ref{fig:Jup_McMZ} is a modified version of Fig.~1 in FN10 and shows model 
results for the core mass and envelope mass of metals of Jupiter predicted by different 
equations of state. These are DFT-MD\footnote{Density Functional Theory for the electrons
and Molecular Dynamics simulations for the ions in the warm dense matter regime}
\citep{Militzer+08}, LM-REOS based on FT-DFT-MD\footnote{In addition to DFT-MD, 
inclusion of finite temperature effects on the electronic subsystem, which 
affect the forces onto the ions} with different EOS representing metals, namely 
He4 and H$_2$O-REOS ~\citep{N-Jupiter+08}, SCvH-i and Sesame-p~\citep{Gui99,SauGui04}, 
Sesame-92 (the original Sesame-EOS for H and He~\citep{SESAME} together with H$_2$O-REOS), 
and Sesame-K04 (a revised H and He Sesame-EOS with minor constituent EOSs from the 
Panda-Code; see \citealt{KTS04-1}). The Jupiter models are published in the same references 
apart from the Sesame-92 based models, which were calculated in \citet{N-PhD-09}. 
For each particular equation of state, numerous solutions are found due to 
uncertainties in the constraints applied. Obviously, a large uncertainty in Jupiter's 
core mass ($0-18\ME$) and envelope metallicity ($0-38\ME$) arises from the 
broad variety of currently competing hydrogen equations of state. The 
total mass of metals $M_{core}+M_Z$ is better constrained and amounts to $14-38\ME$ 
(more precisely, in the order of the above EOS listing: $14-24$, $20-24$ and $28-32$, 
$17-37$, 33, 38, and $35\ME$). Thus  to our current knowledge $4.4-12\%$ of Jupiter's mass 
are heavy elements, which is an enrichment factor of $3-8$ over solar metallicity. 

Since core mass and envelope mass of metals are determined by the requirement to find 
internal density distributions that reproduce $M_p(R_p)$, $J_2$ and $J_4$, the results 
reflect the compressibility of the H/He subsystem along the quasi-adiabat. It is then 
particularly strange that similar methods used to calculate H and He EOS yield 
contrary results on $M_{core}$ and $M_Z$, as it is the case for DFT-MD and LM-REOS based 
on FT-DFT-MD 
(\citealp[see][for details]{N-Jupiter+08,Holst+08,Kietz+07,French+09}).

\subsection{The freedom to assume a discontinuity in heavy elements}
\label{ssec:Jup_ZZ}

In \sect\ref{ssec:Jup_McMZ}, some of the models were based on the assumption $Z_1=Z_2$, 
while others (i.e. those using SCvH-i-99, Sesame-K03, Sesame-92, LM-REOS) were not. 
Here we show that this assumption is not for every EOS subject to free choice, 
if the model is required to meet $J_2$, $J_4$, and $J_6$. In particular, LM-REOS based 
Jupiter models require $Z_1\ll Z_2$, while DFT-MD based models $Z_1\approx Z_2$.

\begin{figure}[htb]
\plotone{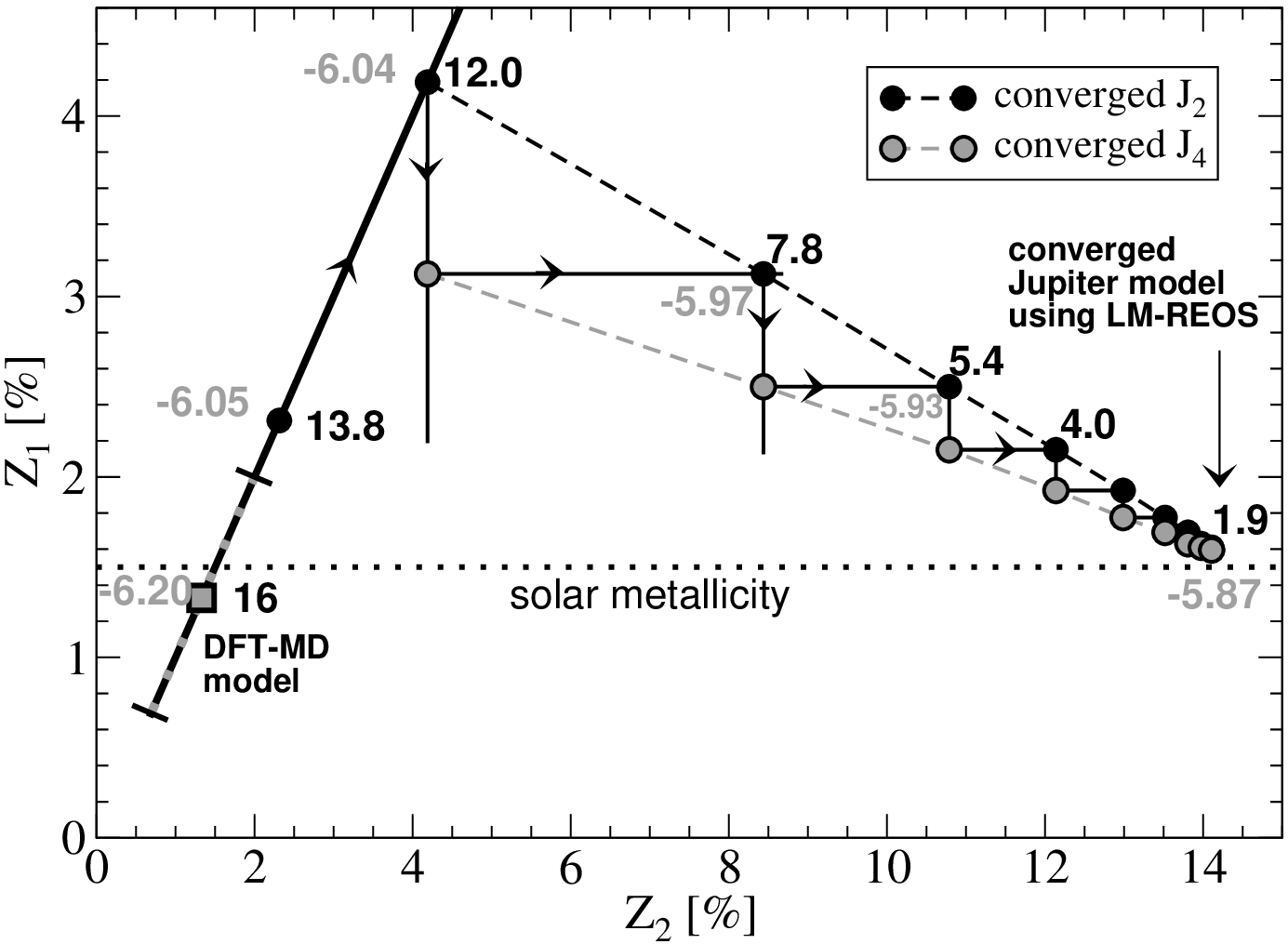}
\caption{\label{fig:Jup_converg}
Converged and intermediate envelope metallicities $Z_1$, $Z_2$ of a three-layer
Jupiter model based on LM-REOS. The converged model matches the most recent 
value $J_4/10^{-4}$ value (\emph{grey numbers}) of $-5.87$ \citep{Jacobson03}, 
has $M_{core}=1.9\ME$ (\emph{black numbers}) and $Z_1=1.6\%$. This model 
was obtained by an interative procedure that produced intermediate solutions 
along the \emph{solid black line}. Starting from $Z_1=Z_2=1\%$, the envelope 
metallicity is increased along the $Z_1=Z_2$ line of homogeneous metallicity 
(\emph{thick black solid}) until $J_2$ is matched. Such a model with 
LM-REOS would have $M_{core}=12.0\ME$ and $J_4/10^{-4}=-6.04$. Then iteratively 
$J_4$ and $J_2$ are fitted by allowing for $Z_1\not=Z_2$ as indicated respectively 
by \emph{grey circles} and \emph{black circles} until outer and inner envelope 
metallicities are found that match both $J_2$ and $J_4$ (converged model). All 
models along the \emph{dashed black line} match $J_2$, and all models along the 
\emph{dashed grey line} match $J_4$. This procedure cannot be applied to 
bring the $J_4$ value of the DFT-MD based Jupiter model 
(\emph{grey square with error bars}) \citep{Militzer+08} in agreement with the 
observed $J_4$ value, since obviously $Z_1\ll Z_{\odot}$ would rapidly occur.
}
\end{figure}

The convergent procedure used for LM-REOS based models is illustrated in 
\fig\ref{fig:Jup_converg}. Along the \emph{thick solid line}, $Z_1=Z_2$. On that
line, there is one model that matches $J_2$. However, $J_4/10^{-4}=-6.04$ for 
this model is $8\sigma$ afar from the observed $J_4$ value (see figure caption 
for details). Hence the convergent procedure follows the \emph{thin black line} 
in the direction of the \emph{arrows} until both $J_2$ and $J_4$ are met. 
Along that path, the core mass decreases from $12\ME$ down to $1.9\ME$. 
The converged solution has $Z_1\ll Z_2$. This behavior applies to every LM-REOS 
based Jupiter model, and hence the choice $Z_1=Z_2$ cannot be made unless the 
interior does not rotate rigidy but instead the observed $J_4$ 
value is influenced by deep winds~\citep{Militzer+08}. 
 
On the other hand, as \fig\ref{fig:Jup_converg} shows, $Z_1\ll Z_2$ is not a 
favorite option for the DFT-MD based Jupiter model, since that would imply
 $Z_1\ll Z_{\odot}$ or even zero. This would contradict heavy element abundance 
measurements in Jupiter's atmosphere, which indicate $Z_1\geq 2\times Z_{\odot}$.
Therefore, $Z_1\approx Z_2$ is \emph{not} a free choice for the DFT-MD based
model. Consequently, the very different Jupiter core masses obtained using 
these two ab-initio EOS are not primarily due to different assumptions about 
the distribution of heavy elements. In order to resolve this problem, a comparison
of the H/He adiabats is highly desirable. Furthermore, isentropic compression 
experiments in the $0.5-4$~Mbar regime where $J_2$ and $J_4$ are most sensitive 
to metallicty would greatly help to discriminate between competing Jupiter models. 
For the other EOS considered here, both the assumptions $Z_1=Z_2$ or $Z_1\not=Z_2$
give acceptable Jupiter models and hence are a matter of free choice.

\section{Extrasolar planet models}\label{sec:J2k2}

\subsection{$J_2$ and $k_2$}

As we have seen in \sect\ref{sec:Jupiter}, Jupiter models that use one 
particular EOS can have various resulting $\{M_{core}, Z_1, Z_2\}$ triples. 
Without the constraint imposed by $J_4$ however, the variety of solutions 
would be immense, including $M_{core}>18\ME$ and $M_Z>38\ME$. 
For extrasolar planets, currently available gravity data are $M_p$ and 
$R_p$ only. A potentially observational constraint equivalent to $J_2$ 
has been suggested to be the tidal Love number $k_2$ \citep{RW09}, 
which measures the ability of a planet to develop an elliptic deformation 
in response to a tidal perturber such as the close-by parent star. 
In this section, we present implications of knowing a precise $k_2$ value 
of the Neptune-sized extrasolar planet GJ~436b.

\subsection{Core mass of GJ 436b models for given $k_2$}

\begin{figure}[htb]
\plotone{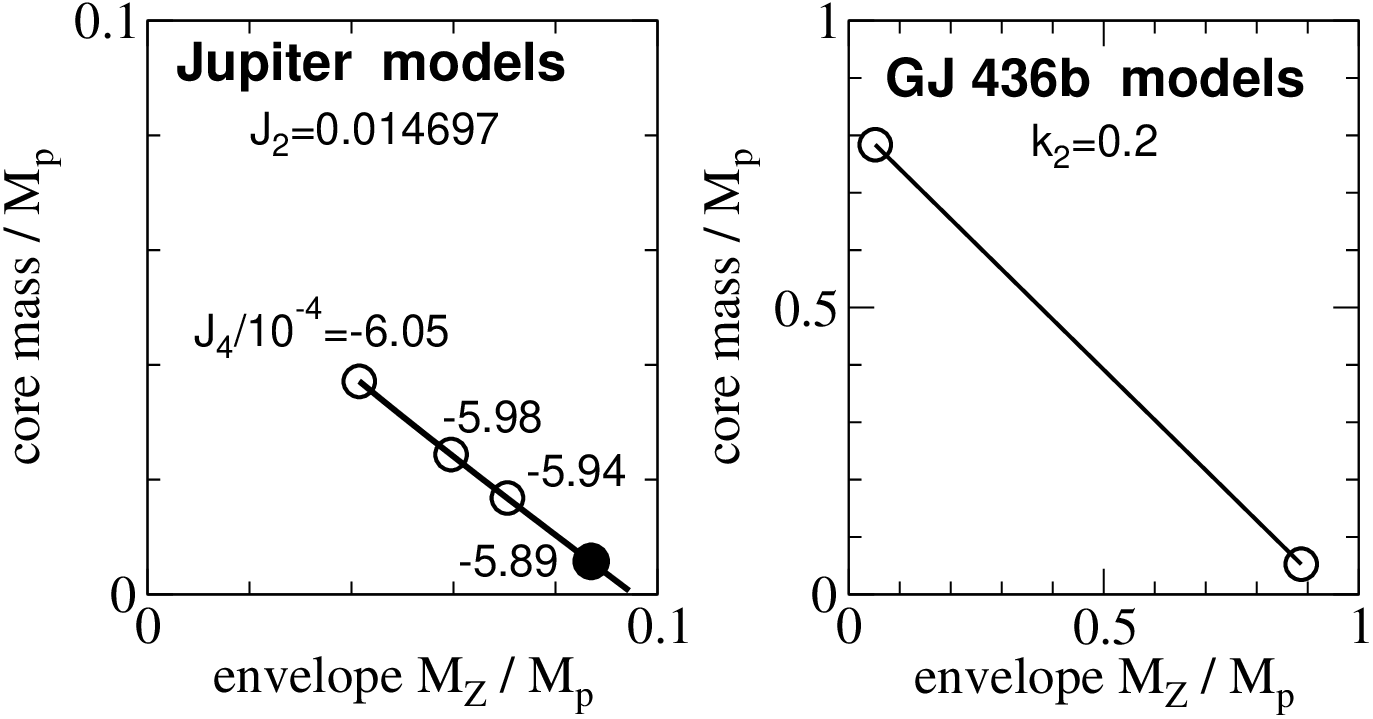}
\caption{\label{fig:J2k2}
Jupiter models matching the measured $J_2$ value, and GJ$\:$436b models matching 
an arbitrary $k_2$ value. Maximum core mass (no water layer) and minimum core mass
(no rock core) solutions are found. Accurate ($\sim 1\%$) knowledge 
of Jupiter's $J_4$ reduces the uncertainties in $M_{core}$ and $M_Z$ (\emph{filled 
circle} solution), while GJ$\:$436b's core mass is undetermined if $k_2=0.2$ and 
$k_4$ remains unknown.}
\end{figure}

About 30 light years away from Earth, the Hot Neptune GJ$\:$436b 
($M_p=23.17\ME$, $R_p=4.22\RE$)~\citep{Torres+08} orbits the M-star GJ$\:$436. 
Placed in a mass-radius diagram, this planet is located close to theoretical 
$M$-$R$ relations of warm water planets. Accordingly, interior structure models assuming an 
iron-silicate core, a water layer, and a H/He envelope allow for a composition 
of 95\% water and 5\% H/He \citep{Figueira+09}, but also for a water-less 
composition. H/He envelope models, whether dry or not, are found to have 
$0.02<k_2<0.2$ \citep{N-GJ436b+10}. We here \emph{assume} $k_2=0.2$ and 
search for the core mass and water content that satisfy this additional 
constraint. We allow water be mixed homogeneously into the H/He envelope 
(two-layer model) or confined to a deep water layer (three-layer models) and 
combinations in between (Neptune-like three-layer models). Figure~\ref{fig:J2k2} 
shows the result in comparison with LM-REOS based Jupiter models forced to meet 
the observed $J_2$ value, but not $J_4$. 

In both cases, the solutions are located along a straight line (solutions for 
different $k_2$ or $J_2$ values would span parallel lines). For 
both planets, and hence in general, the upper limit in core mass is obtained
for homogeneous envelope models (compare \sect\ref{ssec:Jup_ZZ}),
and the upper limit in envelope mass of metals (here water) by zero-mass
core models. In between there are three-layer models with various metallicities
in the two envelopes. As indicated for Jupiter, solutions below the 
\emph{filled circle} are also in agreement with $J_4$.

We conclude that the constraint imposed by $k_2$ or $J_2$ reduces the degeneracy
of models, but further constraints are needed to determine the core mass
of a planet within a few percent of total mass (Jupiter: 6\%). Cooling curve
calculations might help to constrain the composition, since they depend on 
specific heat which significantly varies between relevant bulk materials. 
Unfortunately, evolution calculations are subject to uncertainties
such as tidal interactions \citep{Leconte+10} and stellar irradiation 
and the response of the atmosphere (e.g.;\citealp{Valencia+10}).

\section{Conclusions}

Standard Jupiter interior models predict a core mass of $0-18\ME$, 
an envelope mass of metals of $0-38\ME$, and a total mass of metals of $14-38\ME$.
These uncertainties have increased over the past ten years \citep{Gui99}, 
when they were thought to arise from the possible presence of a plasma phase 
transition in hydrogen. From the behavior of model solutions discussed in this 
article, we conclude an unsufficient understanding of the H/He EOS in the 0.5$-$4 
Mbar pressure regime, where $J_2$ and $J_4$ are most sensitive.
Equivalent to Jupiter models, measuring the $k_2$ value of an extrasolar 
planet would reduce possible solutions to a linear relation between core mass 
and envelope mass of metals, as shown for GJ$\:$436b.

\acknowledgements

The author kindly thanks  R.~Redmer, J.J.~Fortney, T.~Guillot, and D.J.~Stevenson 
for insightful discussions, U.~Kramm for providing the $k_2$ values, and the 
anonymous referee for valuable comments.

\begin{footnotesize}
\bibliographystyle{spr-mp-nameyear-cnd}
\bibliography{nnettelmann-refs}
\end{footnotesize}

\end{document}